\documentclass[11pt,reqno,a4paper,twoside]{article}

\usepackage[english]{babel}
\usepackage{latexsym,amsfonts,amsmath,amsthm,amstext,amscd,amssymb,euscript,wasysym}
\usepackage{rotating}
\usepackage{multirow}
\usepackage{array}
\usepackage{enumitem}
\usepackage{graphicx}
\usepackage{subfigure}
\usepackage{verbatim} %this allows for comment on a whole paragraph
\usepackage{amsbsy} %this gives bold symbols
\usepackage{epstopdf}
\usepackage{wasysym}

\usepackage{leftidx}

\usepackage[usenames,dvipsnames]{pstricks}
\usepackage{epsfig}
\usepackage{pst-grad} % For gradients
\usepackage{pst-plot} % For axes
\makeatletter
\@addtoreset{equation}{section}
\makeatother

\makeatletter
\@addtoreset{enunciato}{section}
\makeatother

\newcounter{enunciato}[section]

\newtheorem{ittheorem}{Theorem}
\newtheorem{itlemma}{Lemma}
\newtheorem{itproposition}{Proposition}
\newtheorem{itdefinition}{Definition}
\newtheorem{itcorollary}{Corollary}

\newenvironment{theorem}{\addtocounter{enunciato}{1}
\begin{ittheorem}}{\end{ittheorem}}

\newenvironment{lemma}{\addtocounter{enunciato}{1}
\begin{itlemma}}{\end{itlemma}}

\newenvironment{proposition}{\addtocounter{enunciato}{1}
\begin{itproposition}}{\end{itproposition}}

\newenvironment{definition}{\addtocounter{enunciato}{1}
\begin{itdefinition}}{\end{itdefinition}}

\newcommand{\be}{\begin{equation}}
\newcommand{\ee}{\end{equation}}

\newcommand{\bt}[1]{\begin{theorem}\label{#1}}
\newcommand{\et}{\end{theorem}}

\newcommand{\bl}[1]{\begin{lemma}\label{#1}}
\newcommand{\el}{\end{lemma}}

\newcommand{\bp}[1]{\begin{proposition}\label{#1}}
\newcommand{\ep}{\end{proposition}}

\newcommand{\bd}[1]{\begin{definition}\label{#1}}
\newcommand{\ed}{\end{definition}}

\newcommand{\bpr}{\begin{proof}}
\newcommand{\epr}{\end{proof}}

\def\Dn{\Delta_n}
\def\Tn{\mathbb{T}_n}
\def\T{\mathbb{T}}
\def\M{\mathcal{M}}
\def\P{\mathbb{P}}
\def\CW{\mathrm{CW}}
\def\iu{\lfloor n u \rfloor}

\renewcommand{\phi}{\varphi}
\renewcommand{\subset}{\subseteq}
\renewcommand{\emptyset}{\varnothing}

\newcommand{\Z}{\mathbb Z}
\newcommand{\R}{\mathbb R}
\newcommand{\N}{\mathbb N}

\newcommand{\cN}{\mathcal N}
\newcommand{\cI}{\mathcal I}
\newcommand{\cL}{\mathcal L}
\newcommand{\cX}{\mathcal X}

\newcommand{\If}[1]{\mathbb{I}_{#1}}

\DeclareMathOperator{\argmin}{argmin}

\begin{document}

\title{Variational description of Gibbs-non-Gibbs\\ 
dynamical transitions for spin-flip systems\\ 
with a Kac-type interaction}

\author{
\renewcommand{\thefootnote}{\arabic{footnote}}
R. Fern\'andez
\footnotemark[1]
\\
\renewcommand{\thefootnote}{\arabic{footnote}}
F. den Hollander
\footnotemark[2]
\\
\renewcommand{\thefootnote}{\arabic{footnote}}
J. Mart\'inez
\footnotemark[3]
}

\footnotetext[1]{
Department of Mathematics, Utrecht University, P.O.\ Box 80010, 3508 TA Utrecht, 
The Netherlands, {\sl R.Fernandez1@uu.nl}
}
\footnotetext[2]{
Mathematical Institute, Leiden University, P.O.\ Box 9512, 2300 RA, Leiden, The 
Netherlands, \newline {\sl denholla@math.leidenuniv.nl}
}
\footnotetext[3]{
Mathematical Institute, Leiden University, P.O.\ Box 9512, 2300 RA, Leiden, The 
Netherlands, \newline {\sl martinez@math.leidenuniv.nl}
}

\maketitle

\begin{abstract}
We continue our study of Gibbs-non-Gibbs dynamical transitions. In the present paper 
we consider a system of Ising spins on a large discrete torus with a Kac-type interaction 
subject to an independent spin-flip dynamics (infinite-temperature Glauber dynamics). 
We show that, in accordance with the program outlined in \cite{vEFedHoRe10}, in the 
thermodynamic limit Gibbs-non-Gibbs dynamical transitions are \emph{equivalent} to 
bifurcations in the set of global minima of the large-deviation rate function for the 
trajectories of the empirical density \emph{conditional} on their endpoint. More precisely, 
the time-evolved measure is non-Gibbs if and only if this set is not a singleton for 
\emph{some} value of the endpoint. A partial description of the possible scenarios of 
bifurcation is given, leading to a characterization of passages from Gibbs to non-Gibbs 
and vice versa, with sharp transition times. 

Our analysis provides a conceptual step-up from our earlier work on Gibbs-non-Gibbs 
dynamical transitions for the Curie-Weiss model, where the mean-field interaction allowed 
us to focus on trajectories of the empirical magnetization rather than the empirical density.  

{\it Acknowledgment.} FdH is supported by ERC Advanced Grant VARIS-267356. JM is supported
by Erasmus Mundus scholarship BAPE-2009-1669. The authors are grateful to A.\ van Enter and 
F.\ Redig for ongoing discussions on non-Gibbsianness, and to T.\ Franco and M.\ Jara for help 
with the hydrodynamic scaling argument in Appendix~\ref{appA}.
\end{abstract}

\newpage

%%%%%%%%%%%%%%%%%% SECTION 1 %%%%%%%%%%%%%%%%%%%

\section{Introduction and main results}
\label{S1}

%%%%%%%%%%%%%%%%%%%%%%

\subsection{Background}
\label{S1.1}

Gibbs-non-Gibbs dynamical transitions are a surprising phenomenon. An initial Gibbsian 
state (e.g.\ a collection of interacting Ising spins) is subjected to a stochastic 
dynamics (e.g.\ a Glauber dynamics) at a temperature that is \emph{different} 
from that of the initial state. For many combinations of initial and dynamical temperature, 
the time-evolved state is observed to become non-Gibbs after a finite time. Such a state 
cannot be described by any absolutely summable Hamiltonian and therefore \emph{lacks a 
well-defined notion of temperature}.   

The phenomenon was originally discovered by van Enter, Fern\'andez, den Hollander
and Redig~\cite{vEFedHoRe02} for \emph{heating dynamics}, in which a low-temperature 
Ising model is subjected to a high-temperature Glauber dynamics. The state remains 
Gibbs for short times, but becomes non-Gibbs after a finite time. Remarkably, heating 
in this case does not lead to a succession of states with increasing temperature, but 
to states where the notion of temperature is \emph{lost altogether}. Moreover, it 
turned out that there is a difference depending on whether the initial Ising model 
has zero or non-zero magnetic field. In the former case, non-Gibbsianness once lost 
is never recovered, while in the latter case Gibbsianness is recovered at a later time. 

This initial work triggered a decade of developments. By now, results are available
for a variety of interacting particle systems, both for \emph{heating dynamics} and 
for \emph{cooling dynamics}, including estimates on transition times and characterizations 
of the so-called \emph{bad configurations} leading to non-Gibbsianness, i.e., the 
discontinuity points of the conditional probabilities. It has become clear that 
Gibbs-non-Gibbs transitions are the rule rather than the exception. For references 
we refer to the recent overview by van Enter~\cite{vE12}.

%%%%%%%%%%%%%%%%%%%%%%%%%%%%%%

\subsection{Motivation and outline}
\label{S1.2}

The ubiquity of the Gibbs-non-Gibbs phenomenon calls for a better understanding of 
its causes and consequences. Historically, non-Gibbsianness is proved by looking at 
the evolving system at two times, the inital time and the final time, and applying 
techniques from equilibrium statistical mechanics. This is an indirect approach 
that does not illuminate the relation between the Gibbs-non-Gibbs phenomenon and 
the dynamical effects responsible for its occurrence. This unsatisfactory situation 
was addressed in Enter, Fern\'andez, den Hollander and Redig~\cite{vEFedHoRe10}, 
where possible dynamical mechanisms were proposed and a \emph{program} was put 
forward to develop a theory of Gibbs-non-Gibbs transitions on \emph{purely dynamical 
grounds}. 

In Fern\'andez, den Hollander and Mart\'inez~\cite{FedHoMa13}, building on earlier
work by K\"ulske and Le Ny~\cite{KuLeNy07} and Ermolaev and K\"ulske~\cite{ErKu10}, 
we showed that this program can be fully carried out for the Curie-Weiss model subject 
to an infinite-temperature dynamics. The goal of the present paper is to extend this 
work away from the mean-field setting by considering a model with a Kac-type interaction, 
i.e., Ising spins with a long-range interaction. Whereas for the Curie-Weiss model the key 
object was the empirical magnetization in the thermodynamic limit, for the Kac model 
the key object is the \emph{empirical density} in the thermodynamic limit, which 
we refer to as the \emph{profile}. Non-Gibbsianness corresponds to a discontinuous 
dependence of the law of the initial profile \emph{conditional} on the final profile. The 
discontinuity points are called \emph{bad profiles} (Definition~\ref{goodpoint} below). 

Dynamically, such discontinuities are expected to arise whenever there is more than 
one trajectory of the profile that is \emph{compatible} with the bad profile at the end. 
Indeed, this expectation is confirmed and exploited in the sequel. The actual conditional 
trajectories are those minimizing the large-deviation rate function on the space of trajectories 
(Propositions~\ref{LDPCo}--\ref{PQLDPS} below), in the spirit of what is behind hydrodynamic 
scaling. The time-evolved measure is Gibbs whenever there is a single minimizing trajectory 
for every final profile, in which case the so-called specification kernel can be computed 
explicitly (Theorem~\ref{Thetheorem} below). In contrast, if there are multiple optimal 
trajectories, then the choice of trajectory can be decided by an infinitesimal perturbation 
of the final profile, and the time-evolved measure is non-Gibbs (Theorem~\ref{nGBi} 
below). 

The rate function for the Kac model contains an action integral whose Lagrangian acts 
on profiles. This setting constitutes a conceptual step-up from what happens for the 
Curie-Weiss model, where the Lagrangian acts on magnetizations and is much easier 
to analyze. However, for infinite-temperature dynamics the Kac Lagrangian can be 
expressed as an integral of the Curie-Weiss Lagrangian with respect to the profile 
(Theorem~\ref{KacCWrel} below). This link allows us to identify the possible scenarios 
of bifurcation (Theorem~\ref{Bifurc} below).

%%%%%%%%%%%%%%%%%%%%%%%

\subsection{Hamiltonian}
\label{S1.3}

Let $\T^d:=\R^d/\Z^d$ be the $d$-dimensional unit torus. For $n \in \N$, let $\Tn^d$ be
the $(1/n)$-discretization of $\T^d$ defined by $\Tn^d:=\Dn^d/n$, with $\Dn^d:=\Z^d/n\Z^d$ 
the discrete torus of size $n$. For $n \in \N$, let $\Omega_n:=\{-1,+1\}^{\Dn^d}$ be the 
set of Ising-spin configurations on $\Dn^d$. The energy of the configuration $\sigma
:=(\sigma(x))_{x \in \Dn^d} \in \Omega_n$ is given by the \emph{Kac-type Hamiltonian}
\be
\label{Hamiltonian}
H^n(\sigma):=-\tfrac{1}{2 n^d} \sum_{x,y \in \Dn^d} 
J\left(\tfrac{x-y}{n}\right)\,\sigma(x)\sigma(y)
-\sum_{x \in \Dn^d} h(\tfrac{x}{n})\,\sigma(x),
\qquad \sigma \in \Omega_n,
\ee
where $J,h \in C(\T^d)$ are continuous functions on $\T^d$, with $J \geq 0$ symmetric
and $J \not\equiv 0$. The Gibbs measure associated with $H^n$ is
\be
\label{Kac}
\mu^n(\sigma):=\frac{e^{-\beta H^n(\sigma)}}{Z^n}, \qquad \sigma \in \Omega_n,
\ee
with $\beta\in [0,\infty)$ the \emph{static} inverse temperature and $Z^n$ the normalizing 
partition sum.

%%%%%%%%%%%%%%%%%%%%%%%%%%%%%%%%%%%%%%%%

\subsection{Gibbs versus non-Gibbs}
\label{S1.4}

For $\Lambda \subseteq \Dn^d$, let $\pi_{\Lambda}^n\colon\,\Omega_n \to \M(\Tn^d)
\subseteq \M(\T^d)$ be the \emph{empirical density} of $\sigma$ inside $\Lambda$ defined 
by
\be
\label{empDensity}
\pi^n_{\Lambda}(\sigma) := \frac{1}{|\Lambda|} 
\sum\limits_{x \in \Lambda} \sigma(x) \delta_{x/n},
\ee
where $\M(\Tn^d)$ and $\M(\T^d)$ denote the set of signed measures on $\Tn^d$, 
respectively, $\T^d$ with total variation norm $\leq 1$ endowed with the weak topology,
and $\delta_{u}$ is the point measure at $u\in\T^d$. Note that $\sigma\in\Omega_n$ 
determines $\pi^n_{\Lambda} \in \M(\Tn^d)$ and vice versa.

Abbreviate \eqref{empDensity} for $\Lambda= \Dn^d$ by $\pi^n$ and for $\Lambda = \Dn^d 
\backslash \{\iu\}$ by $\pi^{u,n}$, $u \in \T^d$, where $\iu$ denotes the component-wise 
lower-integer part of $nu$. The latter is the empirical density \emph{perforated} at $\iu$.
Abbreviate 
\be
\label{Mndefs}
\M^n:=\pi^n(\Omega_n), \qquad \M^{u,n} := \pi^{u,n}(\Omega_n).
\ee 
Note that $\M^n \subseteq \M(\Tn^d)$. Via $\pi^n$, the Gibbs measure $\mu^n$ on 
$\Omega_n$ in \eqref{Kac} induces a probability measure $\check{\mu}^n$ on $\M^n$ 
given by 
\be
\label{mucheck}
\check{\mu}^n = \mu^n \circ (\pi^n)^{-1}.
\ee

Using \eqref{empDensity}, we can rewrite \eqref{Hamiltonian} in the form
\be
\label{Hamiltonianrew}
H^n(\sigma)=-n^d H(\pi^n(\sigma)),
\ee
where in the right-hand side we introduce the notation
\be
\label{Halt}
H(\nu)=\Big\langle \tfrac{1}{2}J\ast\nu + h,\nu \Big\rangle
\ee
\be
\label{inprod}
[f \ast \nu](u) := \int\limits_{\T^d} J(u-u')\,\nu(du'), \qquad 
\langle f,\nu \rangle := \int\limits_{\T^d} f(u)\,\nu(du), \qquad 
f \in C(\T^d),\,\nu \in \M(\T^d).
\ee

Let $\lambda^n := \frac{1}{n^d} \sum_{x\in\Lambda} \delta_{x/n}$. We have $w-\lim_{n \to \infty} 
\lambda^n=\lambda$, where $\lambda$ is the Lebesgue measure on $\T^d$ and $w-\lim$ stands 
for weak convergence. In what follows we will represent limit distributions in $\M(\T^d)$ with a 
Lebesgue density as measures $\alpha \lambda$ with $\alpha \in B$, where 
\be
B \mbox{ is the closed unit ball in } L^\infty(\T^d). 
\ee
We will refer to $\alpha$ as a \emph{profile}.

The definition of Gibbs versus non-Gibbs is the following. Given any sequence
$(\rho^n)_{n\in\N}$ with $\rho^n$ a probability measure on $\Omega_n$ for every 
$n\in\N$, define the single-spin conditional probabilities at site $\iu \in \T^d$ as
\be
\label{gammaun}
\gamma^{u,n} \big(\,\cdot \mid \alpha^u_{n-1} \big)
:= \rho^n \big(\sigma(\iu)= \cdot \mid \pi^{u,n}(\sigma)=\alpha^u_{n-1} \big), 
\qquad \alpha^u_{n-1} \in \M^{u,n}.
\ee

\begin{definition}
\label{goodpoint}
{\rm \bf [Good and bad profiles, Gibbs]}\\
(a) A profile $\alpha \in B$ is called good for $(\rho^n)_{n\in\N}$ if there exists a 
neighborhood $\mathcal{N}_\alpha$ of $\alpha$ in $L^\infty(\T^d)$ such that
\be
\label{specificationlimit}
\gamma^u\big(\,\cdot \mid \tilde{\alpha}\big) 
:= \lim_{n\to\infty} \gamma^{u,n}(\,\cdot \mid \alpha^u_{n-1})
\ee
exists for all $\tilde{\alpha}\in\mathcal{N}_\alpha$ and $u\in\T^d$, for all sequences 
$(\alpha^u_{n-1})_{n\in\N}$ with $\alpha^u_{n-1}\in \M^{u,n}$ for every $n\in\N$ such 
that $w-\lim_{n\to\infty} \alpha^u_{n-1}=\tilde{\alpha}\lambda$, and the limit is independent 
of the choice of $(\alpha^u_{n-1})_{n\in\N}$.\\
(b) A profile $\alpha \in B$ is called bad for $(\rho^n)_{n\in\N}$ if it is not good for 
$(\rho^n)_{n\in\N}$.\\
(c) $(\rho^n)_{n\in\N}$ is called Gibbs if it has no bad profiles in $B$.
\end{definition}

\medskip\noindent
{\bf Remark:} \\
\noindent
(1) Definition~\ref{goodpoint}(a) implies continuity of $\alpha\mapsto\gamma^u(\,\cdot\mid\alpha)$ 
in the $L^\infty(\T^d)$-norm for all $u\in\T^d$ at good profiles. (A proof by contradiction is based on
a diagonal argument.) \\
(2) For $(\mu^n)_{n\in\N}$ with $\mu^n$ defined in (\ref{Hamiltonian}--\ref{Kac}) all profiles 
$\alpha \in B$ are good with
\be
\label{specificationlimitgamma}
\gamma^u(k \mid \alpha) = \frac{\exp[k\beta\{J\ast\alpha+h\}(u)]}
{2\cosh[\beta\{J\ast\alpha+h\}(u)]}, \qquad k \in \{-1,+1\},\,\alpha\in B,\, u \in \T^d.
\ee
(The factor $\tfrac12$ in \eqref{Halt} drops out because every spin is counted
twice in the Hamiltonian but once in the convolution.) In particular, $(\mu^n)_{n\in\N}$ is 
Gibbs in the sense of Definition~\ref{goodpoint}(c).\\
(3) Definition~\ref{goodpoint} assigns the notion of Gibbs to a sequence of probability 
measures that live on different spaces. It is different from the classical notion of Gibbs 
based on the Dobrushin-Lanford-Ruelle condition, which is used to define Gibbs 
measures on infinite lattices. Nonetheless, the quantity in \eqref{specificationlimitgamma} 
can be viewed as some sort of \emph{specification kernel}.\\
(4) Definition~\ref{goodpoint} does not consider sequences $(\alpha^u_{n-1})_{n\in\N}$
whose weak limit is singular with respect to $\lambda$. In Proposition~\ref{LDPCo} below 
we will see that in the thermodynamic limit we can ignore trajectories that do not lie in the 
set $\{\alpha\lambda\colon\,\alpha \in B\}$ because they are too costly.

%%%%%%%%%%%%%%%%%%%%%%%%%%%%%%%%%

\subsection{Stochastic dynamics}
\label{S1.5}

For fixed $n$, we let the spin configuration evolve according to a Glauber dynamics with 
generator $L_n$ given by
\be
\label{generator}
(L_n f)(\sigma) := \sum\limits_{x \in \Dn^d} c_n(x,\sigma)\, 
[f(\sigma^x)-f(\sigma)], \qquad f\colon\,\Omega_n \rightarrow \R,
\ee
where the spin-flip rate takes the form
\be
\label{Metropmod}
c_n(x,\sigma) := \frac{\exp[-\tfrac{\beta'}{2}\{H^n(\sigma^x)-H^n(\sigma)\}]}
{2 \cosh[\tfrac{\beta'}{2}\{H^n(\sigma^x)-H^n(\sigma)\}]}
\ee
with $\sigma^x$ the configuration obtained from $\sigma$ by flipping the spin at site 
$x$, and $\beta' \in [0,\infty)$ the \emph{dynamical} inverse temperature. We write 
$(\sigma_s)_{s \geq 0}$ to denote the trajectory of the spin configuration, which lives 
on $D_{[0,\infty)}(\Omega_n)$, the space of c\`adl\`ag paths on $\Omega_n$ endowed 
with the Skorohod topology.
 
Abbreviate $\pi^n_s:=\pi^n(\sigma_s)$, and let $\bar{\pi}^n=(\pi^n_s)_{s \geq 0}$ denote the 
trajectory of the empirical density under the Glauber dynamics. For a given probability 
measure $\check{\rho}^n_0 $ on $\M^n$ we define 
\be
P^n_{\check{\rho}^n_0} :=  \mbox{ law of } (\pi^n_s)_{s \geq 0} \mbox{ conditional on }
\pi_0^n \mbox{ being drawn according to } \check{\rho}^n_0,
\ee
which lives on $D_{[0,\infty)}(\M^n)$, the space of c\`adl\`ag paths on $\M^n$ 
endowed with the Skorohod topology.

%%%%%%%%%%%%%%%%%%%%%%%%%%%%%%%%%%%%%

\subsection{Large deviation principles}
\label{S1.6}

For $t \geq 0$, we say that $\phi = (\phi_s)_{s \in [0,t]} \in C_{[0,t]}(B)$ is 
absolutely continuous in time when
\begin{equation}
\label{propertyD}
\exists\,\dot{\phi} = (\dot{\phi}_s)_{s \in [0,t]}  \in L^1_{[0,t]}(\T^d)\colon\,\quad
\phi_s(u)-\phi_0(u) = \int_0^s \dot{\phi}_r(u)\,dr \quad \forall\,s  \in [0,t],\,\lambda-a.e.\ u.
\end{equation}

Let us recall that a family of probability measures $(\nu^n)_{n\in\N}$ on a Polish space
$\cX$ satisfies a \emph{large deviation principle} (LDP) with rate $n$ and rate function 
$I$ when $I\colon\,\cX\to [0,\infty]$ has compact level sets, is not identically infinite, and 
\begin{equation}
\begin{aligned}
\liminf_{n\to\infty} \frac1n \log \nu^n(O) 
&\geq & -\inf_{x \in O} I(x), \qquad &O \subset \cX \mbox{ open},\\
\limsup_{n\to\infty} \frac1n \log \nu^n(C) 
&\le & -\inf_{x\in C} I(x), \qquad &C \subset \cX \mbox{ closed}.
\end{aligned}
\end{equation}
(See Dembo and Zeitouni~\cite[Section 1.2]{DeZe10}.) The following LDPs can be found in 
Comets~\cite{Co87}.  

\begin{proposition}
\label{LDPCo}
(i) {\rm {\bf [LDP for initial Gibbs measure]}} 
$(\check{\mu}^n)_{n \in \N}$ satisfies the LDP on $\M(\T^d)$ with rate $n^d$ and rate function 
$I_S-\inf_{\M(\T^d)} I_S$ given by
\be
\label{Srf}
I_S(\nu) :=
\begin{cases}
- \beta \big\langle \tfrac{1}{2} J\ast \alpha + h,\alpha\lambda \big\rangle 
+ \langle\Phi \circ \alpha,\lambda\rangle, 
&\text{if } \nu = \alpha\lambda \mbox{ with } \alpha \in B,\\
\infty, &\text{otherwise},
\end{cases}
\ee
where $\Phi$ is the relative entropy
\be
\label{entropy}
\Phi(m):=\tfrac{1+m}{2}\,\log(1+m)+\tfrac{1-m}{2}\,\log(1-m), \qquad m \in [-1,+1].
\ee
(ii) {\rm {\bf [Dynamical LDP for deterministic initial law]}} 
Let $t \geq 0$ and $\alpha \in C(\T^d)$, and let $({\phi}^n_0)_{n\in\N}$ be any sequence 
with ${\phi}^n_0 \in \M^n$ for every $n\in\N$ such that $w-\lim_{n \to \infty} {\phi}^n_0 
= \alpha \lambda$. Then 
\be
\left(P^n_{\delta_{{\phi}^n_0}}\right)_{n \in \N} \mbox{ restricted to } [0,t]
\ee 
satisfies the LDP on $D_{[0,t]}(\M(\T^d))$ with rate $n^d$ and rate function 
$I_D^t-\inf_{D_{[0,t]}(\M(\T^d))} I_D^t$ given by
\be
\label{Drf}
I_D^t(\psi):=
\begin{cases}
\int_0^t \cL\big(\phi_s,\dot{\phi}_s\big)\,ds, 
&\text{if } \psi=\phi \lambda, \text{ with } \phi 
\text{ satisfying property \eqref{propertyD} and } \phi_0 \equiv \alpha,\\
\infty, & \text{otherwise},
\end{cases}
\ee
where
\be
\label{action}
\cL(p,q) := \int_{\T^d} L[p(u),q(u)]\,du, \qquad p \in B,\,q \in L^1(\T^d),
\ee
with
\be
\label{Lagrangian}
\begin{aligned}
L[p(u),q(u)] 
&= \tfrac{q(u)}{2} \log \Bigg[ \frac{\frac{q(u)}{2} 
+ \sqrt{1-p(u)^2+\big[\frac{q(u)}{2}\big]^2}}{1-p(u)} \Bigg]
-\tfrac{q(u)}{2}\,\big[\beta'(J\ast p + h)\big](u)\\
&\qquad\qquad+  \Bigg\{
- \sqrt{1-p(u)^2+\left[\tfrac{q(u)}{2}\right]^2}\\
&\qquad\qquad\qquad\qquad 
+ \cosh\big[\beta'(J \ast p + h)\big](u) - p(u)\sinh\big[\beta'(J \ast p + h)\big](u)
\Bigg\}.
\end{aligned}
\ee
\end{proposition}

\noindent 
Note that \eqref{Lagrangian} simplifies considerably when $\beta'=0$ (independent 
spin-flip dynamics). 

To ease notation, we write $I_S(\alpha)$ instead of $I_S(\nu)$ when $\nu=\alpha \lambda$ 
with $\alpha \in B$, and $I^t_D(\phi)$ instead of $I^t_D(\psi)$ when $\psi=\phi\lambda$ with 
$\phi \in C_{[0,t]}(B)$, i.e., we henceforth suppress the reference measure $\lambda$
from the notation.

Let $P^n=P^n_{\check{\mu}^n}$. Define
\be
Q^n_{t,\alpha'}(\cdot) := P^n \big( (\pi^n_s)_{s \in [0,t]}  \in \cdot \mid \pi^n_t = \alpha'_n \big), 
\qquad t \geq 0,\ \alpha' \in B,
\ee
with $\alpha'_n \in \M^n$ the element closest to $\alpha' \in B$ in any metric that metrizes 
the weak topology. The following LDPs are key to our analysis. In what follows we write 
$f \equiv g$ when $f(u)=g(u)$ for all $u\in \T^d$. 

\begin{proposition}
\label{PQLDPS}
{\rm {\bf [Dynamical LDP for Gibbs initial law]}}\\ 
(i) For every $t \geq 0$, $(P^n)_{n\in\N}$ satisfies the LDP on $D_{[0,t]}(\M(\T^d))$ with 
rate $n^d$ and rate function $I^t-\inf_{D_{[0,T]}(\M(\T^d))} I^t$ given by
\be
\label{SDrf}
I^t(\phi) := I_S(\phi_0) + I_D^t(\phi).
\ee
(ii) For every $t \geq 0$ and $\alpha' \in B$, $(Q^n_{t,\alpha'})_{n \in \N}$ satisfies the LDP 
on $D_{[0,t]}(\M(\T^d))$ with rate $n^d$ and rate function $I^{t,\alpha'}-\inf_{D_{[0,t](\M(\T^d))}} 
I^{t,\alpha'}$ given by
\be
\label{variational}
I^{t,\alpha'}(\phi)
:= \left\{\begin{array}{ll}
I^t(\phi), &\text{if } \phi_t \equiv \alpha',\\
\infty, &\text{otherwise}.
\end{array}
\right.
\ee
\end{proposition}

\noindent
The proof of Proposition~\ref{PQLDPS} is given in Appendix~\ref{appA} and is based 
on large deviation techniques coming from hydrodynamic scaling. A somewhat delicate 
issue is the fact that we cannot use Proposition~\ref{LDPCo}(ii) because this has a 
deterministic initial condition, while in Proposition~\ref{PQLDPS}(i) the initial condition
is random.

Note that, by \eqref{Srf}, \eqref{Drf} and (\ref{SDrf}--\ref{variational}), 
\be
\label{cvp}
\inf_{\phi \in D_{[0,t]}(\M(\T^d))} I^{t,\alpha'}(\phi) 
= \inf_{\alpha \in B}\,\, 
\inf_{\substack{\phi \in C_{[0,t]}(B)\colon \\ 
\phi_0 \equiv \alpha,\,\phi_t \equiv \alpha'}} 
I^t(\phi)
= \inf_{\substack{\phi \in C_{[0,t]}(B)\colon \\ \phi_t \equiv \alpha'}} 
I^t(\phi).
\ee

%%%%%%%%%%%%%%%%%%%%%%%%%%%%%%%%%%%%%%%%

\subsection{Link to the specification kernel}
\label{S1.7}

Henceforth we only consider trajectories $\phi \in C_{[0,t]}(B)$ satisfying \eqref{propertyD},
because the rate functions are infinite otherwise. The following theorem provides the 
fundamental link between the specification kernel in \eqref{specificationlimit} and the 
minimizer of \eqref{cvp} when it is \emph{unique}.

\begin{theorem}
\label{Thetheorem}
{\rm {\bf [Specification kernel in absence of bifurcation]}} 
Fix $t \geq 0$ and $\alpha' \in B$. Suppose that \eqref{cvp} has a unique minimizing 
path $\hat{\phi}^{t,\alpha'}=(\hat{\phi}_s^{t,\alpha'})_{s \in [0,t]}$. Then the 
specification kernel at time $t$ equals
\be
\label{minvsspecif}
\gamma^u_t(k' \mid \alpha') 
:= \frac{\sum\limits_{k\in\{-1,+1\}} 
\exp\big[k\beta\{J \ast \hat{\phi}_0^{t,\alpha'}+h\}(u)\big]\, p^{u,t,\alpha'}_t(k,k')}
{\sum\limits_{j,j'\in\{-1,+1\}} \exp\big[j\beta\{J \ast \hat{\phi}_0^{t,\alpha'}+h\}(u)\big]\, 
p^{u,t,\alpha'}_t(j,j')}, \qquad 
k' \in \{-1,+1\}, \ u \in \T^d,
\ee
where $p^{u,t,\alpha'}_t(j,j')$ is the probability to go from $j$ at time $0$ to $j'$ at time 
$t$ in the time-inhomogene\-ous Markov process on $\{-1,+1\}$ with generator 
$L^{u,t,\alpha'}_s$ at time $s \in [0,t]$ given by
\be
\label{timedependentgenerator}
\begin{aligned}
&(L^{u,t,\alpha'}_s f)(k) 
= \frac{\exp\big[k\beta'\{J \ast \hat{\phi}_s^{t,\alpha'} + h\}(u)\big]}
{2\cosh\big[\beta'\{J \ast \hat{\phi}_s^{t,\alpha'} + h\}(u)\big]}\,[f(-k)-f(k)],\\
&k \in \{-1,+1\},\, f\colon\,\{-1,+1\} \to \R,\,u \in \T^d,\,s \in [0,t].
\end{aligned}
\ee
\end{theorem}

\noindent
{\bf Remark:}
Note that for $\beta'=0$ (independent spin-flip dynamics) the right-hand side of 
\eqref{timedependentgenerator} simplifies to $\tfrac12[f(-k)-f(k)]$ and that, 
consequently, the right-hand side of \eqref{minvsspecif} depends on the optimal trajectory 
$\hat{\phi}^{t,\alpha'}$ only via its initial value $\hat{\phi}^{t,\alpha'}_0$, and takes the form
\begin{equation}
\label{spec-initialm}
\gamma^u_t(k' \mid \alpha')
= \Gamma_t\big(k',\beta\{J \ast \hat{\phi}^{t,\alpha'}_0+h\}(u)\big)
\end{equation}
for some $\Gamma_t\colon\,\{-1,+1\} \times \R \to [0,1]$, with the property that $m \mapsto 
\Gamma_t(k',m)$ is continuous, strictly increasing for $k'=+1$ and strictly decreasing 
for $k'=-1$.

%%%%%%%%%%%%%%%%%%%%%%%

\subsection{Reduction: critical trajectories}
\label{S1.8}

In what follows we restrict ourselves to the case of infinite-temperature dynamics,
i.e., $\beta'=0$. Let
\be
\label{argmin}
\begin{aligned}
\hat{\phi}^{\alpha;t,\alpha'} &:= \argmin_{\substack{\phi \in C_{[0,t]}(B)\colon\\
\phi_0 \equiv \alpha,\,\phi_t \equiv \alpha'}} I^t(\phi),\\
C_{t,\alpha'}(\alpha) &:=  I^t(\hat{\phi}^{\alpha;t,\alpha'}).
\end{aligned}
\ee

\noindent
{\bf Remark:} Note that
\be
\label{simpler}
\inf_{\alpha \in B} C_{t,\alpha'}(\alpha)
=
\inf_{\substack{\phi \in C_{[0,t]}(B)\colon \\ \phi_t \equiv \alpha'}} 
I^t(\phi).
\ee

The following theorem says that $\hat{\phi}^{\alpha;t,\alpha'}$ is unique for every 
$t \geq 0$ and $\alpha,\alpha' \in B$, and can be computed because the Kac model 
can be linked to the Curie-Weiss model treated in Fern\'andez, den Hollander and 
Mart\'{i}nez~\cite{FedHoMa13}. (In the notation of that paper $\beta$ is absorbed into 
$J,h$.) 

\begin{theorem}
\label{KacCWrel}
{\rm {\bf [Critical trajectories]}}
Let $\beta'=0$. For every $t \geq 0$ and $\alpha,\alpha'\in B$, 
\be
\label{CWchoice}
\hat{\phi}^{\alpha; t,\alpha'}_s(u) 
= \hat{\phi}^{\CW;\alpha(u)}_{t,\alpha'(u)}(s), 
\qquad u\in \T^d,\,s \in [0,t],
\ee
where $\hat{\phi}_{t,m'}^{\CW;m}(s)$, $s \in [0,t]$, is the unique trajectory in $[-1,+1]$ 
between magnetization $m$ at time $0$ and magnetization $m'$ at time $t$ for the 
Curie-Weiss model. Accordingly (see {\rm (\ref{Drf}--\ref{Lagrangian})} and 
{\rm \ref{SDrf}--\ref{variational})}),
\be
\label{simplificationcost}
C_{t,\alpha'}(\alpha) = I_S(\alpha) + \int_{\T^d} du \int_0^t ds\,\,
L^\CW\left[\hat{\phi}^{\CW;\alpha(u)}_{t,\alpha'(u)}(s),
\dot{\hat{\phi}}^{\CW;\alpha(u)}_{t,\alpha'(u)}(s)\right],
\ee
where $L^\CW$ is the Lagrangian of the Curie-Weiss model. The critical points of 
\eqref{simplificationcost} (i.e., the local minima and the local maxima) satisfy the 
functional equation
\be
\label{Pequation}
\sinh[2\beta(J\ast\alpha+h)](u)-\alpha(u)\cosh[2\beta(J\ast\alpha+h)](u)
= \frac{\alpha(u)}{\tanh(2t)}-\frac{\alpha'(u)}{\sinh(2t)}
\quad \text{a.e. } u \in \T^d.
\ee
\end{theorem}

\noindent
In Theorem~\ref{KacCWrel}, the Lagrangian of the Curie-Weiss model is given by 
\be
\label{CWLagrangian}
L^{\CW}(m,\dot{m}) := -\tfrac12 \sqrt{4 \left(1-m^2\right)+\dot{m}^2}
+\tfrac12 \dot{m} \log 
\left(\frac{\sqrt{4 \left(1-m^2\right)+\dot{m}^2}+\dot{m}}{2(1-m)}\right)+1,
\ee
which is the same as \eqref{Lagrangian} with $\beta'=0$, $p(\cdot)=m$ and $q(\cdot)=\dot{m}$, 
and the unique trajectory is given by
\begin{equation}
\label{trajectory}
\hat{\phi}_{t,m'}^{\CW;m}(s) := \frac{1}{\sinh(2t)} \Big\{m \sinh(2(t-s)) + m' \sinh(2s) \Big\},
\qquad 0 \leq s \leq t.
\end{equation}
(See \cite[Eqs.\ (1.16) and (1.28)]{FedHoMa13}.) The intuition behind Theorem~\ref{KacCWrel} 
is that the dynamics has no spatial interaction. Consequently, we may think of $\alpha(u)$ and 
$\alpha'(u)$ as the local initial and final magnetization near $u$, and thereby reduce the 
minimization problem in \eqref{variational} to that of the Curie-Weiss model.

With the help of Theorem~\ref{KacCWrel} we are able to prove the equivalence of non-Gibbs 
and \emph{bifurcation}, the latter meaning that \eqref{cvp} has more than one global minimizer. 
This is in accordance with the program outlined in van Enter, Fern\'andez, den Hollander and 
Redig~\cite{vEFedHoRe10}. 

\begin{theorem}
\label{nGBi}
{\rm {\bf [Equivalence of non-Gibbsianness and bifurcation]}}
Let $\beta'=0$. For every $t \geq 0$, $\tilde{\alpha}' \mapsto \gamma^u_t(~\cdot \mid 
\tilde{\alpha}')$ is continuous at $\alpha' \in B$ for all $u \in \T^d$ if and only if 
$\inf_{\phi \in C_{[0,t]}(B)\colon\,\phi_t \equiv \alpha'} I^t(\phi)$ has a unique 
minimizing path. 
\end{theorem}

\noindent
Thus, non-Gibbsianness is equivalent to the occurrence of \emph{more than one} possible 
history for the \emph{same} $\alpha'$. 

We expect Theorem~\ref{nGBi} to hold for $\beta'>0$ as well, but the present paper deals with 
$\beta'=0$ only.

%%%%%%%%%%%%%%%%%%%%%%%%

\subsection{Bifurcation analysis}
\label{S1.9}

In this section we study for which choice of $J,h,\beta$ and $t,\alpha'$ the variational formula 
in the right-hand side of \eqref{cvp} has a unique global minimizer or has multiple global 
minimizers. According to Definition~\ref{goodpoint} and Theorem~\ref{nGBi}, this distinction
classifies Gibbsianness versus non-Gibbsianness.   

\begin{theorem}
\label{Bifurc}
Let $\beta'=0$ and $\langle J \rangle := \int_{\T^d} J(u)du$.\\
(i) {\bf [Short-time Gibbsianness]} 
There exists a $t_0=t_0(J,h) \in (0,\infty)$ such that \eqref{cvp} has a unique global 
minimizer $\hat\phi^{t,\alpha'}$ for all $0 \leq t \leq t_0$ and all $\alpha'\in B$.\\
(ii) {\bf [Mean-field behaviour]} 
If $h \equiv c \in [0,\infty)$ and $\alpha'\equiv c'\in [-1,+1]$, then the bifurcation 
behaviour is the same as for the Curie-Weiss model with parameters $(J^\CW,h^\CW) = 
(\beta \langle J \rangle,\beta c)$ and final magnetization $c'$:
\begin{center}
\begin{tabular}[c]{|l|c|c|}
\hline
$J^\CW$  & $h^\CW=0$ & $h^\CW>0$ \\
\hline
$(0,1]$ & \multicolumn{2}{l|}{\qquad \qquad \qquad \qquad No bad $c'$ for all $t \geq 0$}\\
\hline
$(1,\tfrac{3}{2}]$ 
& 
{
\begin{pspicture}(0,-0.568125)(2.8709376,0.568125)
\psline[linewidth=0.02cm](0.4609375,0.0296875)(2.8609376,0.0296875)
\psdots[dotsize=0.12](0.4609375,0.0296875)
\usefont{T1}{ptm}{m}{n}
\rput(0.84234375,0.3796875){$\emptyset$}
\psdots[dotsize=0.12](1.4209375,0.0296875)
\usefont{T1}{ptm}{m}{n}
\rput(2.2,0.3796875){$\{0\}$}
\usefont{T1}{ptm}{m}{n}
\rput(0.47234374,-0.3403125){$0$}
\usefont{T1}{ptm}{m}{n}
\rput(1.5423437,-0.3403125){$\Psi_c$}
\end{pspicture} 
}
&
{
\begin{pspicture}(0,-0.568125)(5,0.8)
\psline[linewidth=0.02cm](0.4609375,0.0296875)(5,0.0296875)
\psdots[dotsize=0.12](0.4609375,0.0296875)
\usefont{T1}{ptm}{m}{n}
\rput(0.47234374,-0.3403125){$0$}
\usefont{T1}{ptm}{m}{n}
\rput(0.84234375,0.3796875){$\emptyset$}
\psdots[dotsize=0.12](1.4209375,0.0296875)
\usefont{T1}{ptm}{m}{n}
\rput(1.3,-0.3403125){$\Psi_U$}
\usefont{T1}{ptm}{m}{n}
\rput(3,0.3796875){$\{c'\}$}\
\rput(3,-0.2){\small $[-1,U_B]$}
\psdots[dotsize=0.12](4.5,0.0296875)
\usefont{T1}{ptm}{m}{n}
\rput(4.6,-0.3403125){$\Psi_*$}
\rput(5,0.3796875){$\emptyset$}
\end{pspicture} 
}
\\
\hline
\multirow{2}{*}{$(\tfrac{3}{2},\infty)$}
&
{
\begin{pspicture}(0,-0.568125)(4.1,0.6)
\psline[linewidth=0.02cm](0.4609375,0.0296875)(4,0.0296875)
\psdots[dotsize=0.12](0.4609375,0.0296875)
\usefont{T1}{ptm}{m}{n}
\rput(0.47234374,-0.3403125){$0$}
\usefont{T1}{ptm}{m}{n}
\rput(0.84234375,0.3796875){$\emptyset$}
\psdots[dotsize=0.12](1.4,0.0296875)
\usefont{T1}{ptm}{m}{n}
\rput(1.5,-0.3403125){$\Psi_U$}
\usefont{T1}{ptm}{m}{n}
\rput(2.5,0.3796875){$\{\pm c'\}$}\
\rput(2.6,-0.2){\small $[-U_B,U_B]$}
\psdots[dotsize=0.12](3.6,0.0296875)
\usefont{T1}{ptm}{m}{n}
\rput(3.7,-0.3403125){$\Psi_c$}
\rput(4.1,0.3796875){$\{0\}$}
\end{pspicture} 
}
&
{
\begin{pspicture}(-0.2,-0.568125)(8.3,0.8)
\psline[linewidth=0.02cm](0.4609375,0.0296875)(8.1,0.0296875)
\psdots[dotsize=0.12](0.4609375,0.0296875)
\usefont{T1}{ptm}{m}{n}
\rput(0.1,0.6){\footnotesize $h<h^*$}
\rput(0.47234374,-0.3403125){$0$}
\usefont{T1}{ptm}{m}{n}
\rput(0.84234375,0.3796875){$\emptyset$}
\psdots[dotsize=0.12](1.4,0.0296875)
\usefont{T1}{ptm}{m}{n}
\rput(1.5,-0.3403125){$\Psi_U$}
\usefont{T1}{ptm}{m}{n}
\rput(2.5,0.3796875){$\{c'\}$}\
\rput(2.6,-0.2){\small $[M_B,U_B)$}
\psdots[dotsize=0.12](3.6,0.0296875)
\usefont{T1}{ptm}{m}{n}
\rput(3.7,-0.3403125){$\Psi_L$}
\usefont{T1}{ptm}{m}{n}
\rput(4.5,0.3796875){$\{c'_{1},c'_2\}$}\
\rput(4.7,-0.2){{\small $(L_B,M_B)$}}
\psdots[dotsize=0.12](5.8,0.0296875)
\usefont{T1}{ptm}{m}{n}
\rput(5.9,-0.3403125){$\Psi_T$}
\usefont{T1}{ptm}{m}{n}
\rput(6.5,0.3796875){$\{c'\}$}\
\rput(6.9,-0.2){\small $[-1,M_T]$}
\psdots[dotsize=0.12](7.9,0.0296875)
\usefont{T1}{ptm}{m}{n}
\rput(8,-0.3403125){$\Psi_*$}
\rput(8.1,0.3796875){$\emptyset$}
\end{pspicture} 
}
\\
& &
{
\begin{pspicture}(-0.2,-0.568125)(8.3,0.8)
\rput(0.1,0.6){\footnotesize $h\geq h^*$}
\psline[linewidth=0.02cm](0.4609375,0.0296875)(5,0.0296875)
\psdots[dotsize=0.12](0.4609375,0.0296875)
\usefont{T1}{ptm}{m}{n}
\rput(0.47234374,-0.3403125){$0$}
\usefont{T1}{ptm}{m}{n}
\rput(0.84234375,0.3796875){$\emptyset$}
\psdots[dotsize=0.12](1.4209375,0.0296875)
\usefont{T1}{ptm}{m}{n}
\rput(1.3,-0.3403125){$\Psi_U$}
\usefont{T1}{ptm}{m}{n}
\rput(3,0.3796875){$\{c'\}$}\
\rput(3,-0.2){\small $[-1,U_B]$}
\psdots[dotsize=0.12](4.5,0.0296875)
\usefont{T1}{ptm}{m}{n}
\rput(4.6,-0.3403125){$\Psi_*$}
\rput(5,0.3796875){$\emptyset$}
\end{pspicture} 
}
\\
\hline
\end{tabular}
\end{center}
\end{theorem}

\medskip\noindent
The above table summarizes the results for the Curie-Weiss model studied in \cite{FedHoMa13}.
The center line represents the time axis. 
In each figure, the symbols on top indicate the set of bad magnetizations (which for the 
Kac-model correspond to bad constant profiles), the intervals below indicate in which range 
the bad magnetizations occur. For further details, in particular, a definition of the times 
$\Psi_U, \Psi_*, \Psi_c, \Psi_L, \Psi_T$ and the magnetizations $U_B, M_B, L_B, M_T$, 
see \cite[Section 1.5.5]{FedHoMa13}.

\noindent
{\bf Remarks:}\\
(1) The existence of a solution of \eqref{cvp} is guaranteed by the lower semi-continuity 
of $\alpha \mapsto C_{t,\alpha'}(\alpha)$, which follows from the lower semi-continuity 
of $\phi_0 \mapsto I_S(\phi_0)$ and $\phi \mapsto I^t_D(\phi)$, together with the fact 
that $w-\lim_{n \to \infty} \alpha_n=\alpha$ implies $w-\lim_{n \to \infty} \hat{\phi}^{\alpha_n;
t,\alpha'} = \hat{\phi}^{\alpha;t,\alpha'}$ in the Skorohod topology by \eqref{trajectory}.\\
(2) The claims in Theorem~\ref{Bifurc}(ii) only concern the case where $\alpha'$ is constant. 
The problem of deciding whether or not there exist multiple global minimizers of \eqref{cvp} 
when $\alpha'$ is not constant presents major difficulties. Similar but easier equations have 
been studied extensively in Comets, Eisele and Schatzman~\cite{CoEiSc86}, De Masi,
Orlandi, Presutti and Triolo~\cite{dMOPT94} and Bates, Chen and Chmaj~\cite{BCC05},
with partial success. An additional complication in our case is that non-constant $\alpha'$ 
brings a non-homogeneous parameter into the problem, which makes the analysis even 
harder.
A full analysis of the global minimizers of \eqref{cvp} as a function of $J$ and $h$ therefore remains a challenge.

%%%%%%%%%%%%%%%%% SECTION 2 %%%%%%%%%%%%%%%%%%%%%%

\section{Proof of Theorems \ref{Thetheorem}--\ref{nGBi}}
\label{S2}

%%%%%%%%%%%%%%%%%%%%%

\subsection{Proof of Theorem \ref{Thetheorem}}
\label{S2.1}

\begin{proof}
Recall that $\pi^{u,n}_t=\pi^{u,n}(\sigma_t)$ defined below \eqref{empDensity} does not 
depend on $\sigma_t(\iu)$. Let $\P^n$ denote the law of $(\sigma_s)_{s \geq 0}$ with 
$\sigma_0$ distributed according to $\mu^n$, and abbreviate $\pi_{<t}^{u,n}:=(\pi^{u,n}_s)_{s \in [0,t)}$
and $\xi^{n-1}_{<t}:=(\xi^{n-1}_s)_{s \in [0,t)}$. Write (recall \eqref{gammaun})
\be
\label{rel1}
\begin{aligned}
&\gamma^{u,n}_t \big(k' \mid \alpha'^u_{n-1}\big)
:= \P^n \Big(\sigma_t(\iu)=k' ~\Big|~ \pi^{u,n}_t = \alpha'^u_{n-1} \Big)\\
&= \int_{D_{[0,t)}(\M^{u,n})} 
\P^n \Big(d\xi_{<t}^{n-1} ~\Big|~\pi^{u,n}_t=\alpha'^u_{n-1}\Big)
\P^n \Big(\sigma_t(\iu)=k' ~\Big|~ \pi_{<t}^{u,n}=  \xi_{<t}^{n-1} \Big)\\
&= \int_{D_{[0,t)}(\M^{u,n})} \P^n \Big(d\xi_{<t}^{n-1} ~\Big|~\pi^{u,n}_t=\alpha'^u_{n-1}\Big)\\
&\times \Bigg\{ 
\sum_{k = \pm 1} \P^n \Big(\sigma_t(\iu)=k'~\Big|~ \sigma_0(\iu)=k,\,\pi_{<t}^{u,n}=\xi_{<t}^{n-1} \Big)
\P^n \Big(\sigma_0(\iu)=k ~\Big|~ \pi_{<t}^{u,n}=\xi_{<t}^{n-1} \Big) 
\Bigg\}.
\end{aligned}
\ee
We proceed by analyzing the three terms under the integral.

\medskip\noindent
(1) The LDP for $(Q^n_{t,\alpha'})_{n\in\N}$ in Proposition~\ref{PQLDPS}(ii), 
together with the assumption that \eqref{cvp} has a unique minimizing path, implies 
\be
\label{rext}
w-\lim_{n\to\infty} \P^n 
\Big(~\cdot ~\Big|~ \pi^{u,n}_t = \alpha'^u_{n-1}\Big) 
= \delta_{\hat{\phi}^{t,\alpha'}_{<t}}(\cdot)
\quad \mbox{ on } \quad D_{[0,t)}(\M(\T^d)).
\ee

\medskip\noindent
(2) Because $(\sigma_s(\iu),\pi^{u,n}_s)_{s \geq 0}$ is Markov, we have  
\be
\label{MP}
\P^n \Big(\sigma_t(\iu)=k' ~\Big|~ \sigma_0(\iu)=k,\,\pi_{<t}^{u,n}=\xi_{<t}^{n-1} \Big)
= p^{\xi^{n-1}_{<t}}_t (k,k'),
\ee
where $p^{\xi^{n-1}_{<t}}_t(k,k')$ is the probability to go from $k$ at time 0 to $k'$ at time $t$ 
in the time-inhomogeneous Markov process on $\{-1,+1\}$ with generator at time $s \in [0,t)$ 
given by \eqref{timedependentgenerator} with $\hat{\phi}_s^{t,\alpha'}$ replaced by $\xi_s^{n-1}$. 
Note that $\xi^{n-1}_{<t} \mapsto p^{\xi^{n-1}_{<t}}_t(k,k')$ is continuous on $D_{[0,t)}(\M^{u,n})$ 
for fixed $k,k',t$ and $u,n$ (recall \eqref{Mndefs}), and that $\lim_{n\to\infty} p^{\xi^{n-1}_{<t}}_t(k,k')
= p^{\hat\phi_{<t}^{t,\alpha'}}_t(k,k')$ for fixed $k,k',t,\alpha'$ when $\lim_{n\to\infty} \xi^{n-1}_{<t}
=\hat\phi_{<t}^{t,\alpha'}$ on $D_{[0,t)}(\M(\T^d))$ (recall \eqref{timedependentgenerator}).

\medskip\noindent
(3) Write
\be
\label{rmid}
\begin{aligned}
&\P^n \Big(
\sigma_0(\iu) = k ~\Big|~ \pi_{<t}^{u,n}=\xi_{<t}^{n-1} \Big) \\
&\qquad\qquad = \left[1+c^{u,n}(\xi_{<t}^{n-1},k) \exp\left(-2\beta k \{\tfrac12 J\ast\xi^{n-1}_0+h\}
\big(\tfrac{\iu}{n}\big)\right)\right]^{-1}
\end{aligned}
\ee
with
\be
c^{u,n}(\xi_{<t}^{n-1},k) := \frac{d\P^{u,n}_{\xi^{n-1}_0,-k}}{d\P^{u,n}_{\xi^{n-1}_0,k}}(\xi^{n-1}_{<t}),
\ee
where
\be
\P^{u,n}_{\xi^{n-1}_0,k}(\cdot) = \P^{u,n}\big(\pi^{u,n}_{<t} \in \cdot \mid  
\,\pi^{u,n}_0=\xi^{n-1}_0, \sigma_0(\iu)=k\big)
\ee
and we use (\ref{Hamiltonian}--\ref{Kac}) to write
\be
\frac{\P^n(\pi^{u,n}_0=\xi^{n-1}_0, \sigma_0(\iu)=-k)}
{\P^n(\pi^{u,n}_0=\xi^{n-1}_0,\, \sigma_0(\iu)=k)}
= \exp\left(-2\beta k \{\tfrac12 J\ast\xi^{n-1}_0+h\}
\big(\tfrac{\iu}{n}\big)\right).
\ee

\medskip\noindent
Finally, note that $\lim_{n\to\infty} c^{u,n}(\xi_{<t}^{n-1},k) = 1$ for fixed $k,t$ and $u$
when $\lim_{n\to\infty} \xi^{n-1}_{<t}=\hat\phi_{<t}^{t,\alpha'}$ on $D_{[0,t)}(\M(\T^d))$. 
Indeed, (\ref{generator}--\ref{Metropmod}) show that in the thermodynamic limit a single 
spin has no effect on the dynamics of the empirical density (Feller property). Combine 
this observation with (\ref{rext}--\ref{rmid}) to get the identity in \eqref{minvsspecif}
(see Yang~\cite{Ya11}).
\end{proof}

%%%%%%%%%%%%%%%%%%%%%%%%%%

\subsection{Proof of Theorem~\ref{KacCWrel}}
\label{S2.2}

\begin{proof}
For $\beta'=0$ (infinite-temperature dynamics), \eqref{Lagrangian} reduces to $\int_{\T^d} 
du\,L^\CW[p(u),q(u)]$ with $L^\CW$ the Curie-Weiss Lagrangian in \eqref{CWLagrangian}. 
Hence, recalling \eqref{variational}, we have
\be
\begin{aligned}
\label{MFreduction}
C_{t,\alpha'}(\alpha) 
&= \inf_{\substack{\phi \in C_{[0,t]}(B)\colon\\
\phi_0 \equiv \alpha,\,\phi_t\equiv \alpha'}} I^t(\phi)\\
&= I_S(\alpha) + 
\inf_{\substack{\phi\in C_{[0,t]}(B)\colon \\ 
\phi_0 \equiv \alpha,\,\phi_t\equiv \alpha'}} I_D^t(\phi) \\
&\geq I_S(\alpha) + \int_{\T^d} du\,
\inf_{\substack{\phi \in C_{[0,t]}(B)\colon\\
\phi_0\equiv\alpha,\,\phi_t\equiv\alpha'}} 
\int_0^t ds\,\,L^\CW\big[\phi_s(u),\dot{\phi}_s(u)]\\
&\geq I_S(\alpha) + \int_{\T^d} du\,
\inf_{\substack{\rho \in C_{[0,t]}([-1,+1])\colon\\
\rho_0=\alpha(u),\,\rho_t=\alpha'(u)}} 
\int_0^t ds\,\,L^\CW\big[\rho_s,\dot{\rho_s}]\\
&= I_S(\alpha) + \int_{\T^d} du
\int_0^t ds\,\,L^\CW\Big[\hat{\phi}_{t,\alpha'(u)}^{\CW;\alpha(u)}(s),
\dot{\hat{\phi}}_{t,\alpha'(u)}^{\CW;\alpha(u)} (s)\Big],
\end{aligned}
\ee
which settles half of \eqref{simplificationcost}. To get equality we pick, 
as in \eqref{CWchoice},
\be
\hat{\phi}^{\alpha;t,\alpha'}_s(u) := \hat{\phi}_{t,\alpha'(u)}^{\CW;\alpha(u)}(s), 
\quad s \in [0,t], \ ,u \in \T^d. 
\ee
Since $(\hat{\phi}^{\alpha;t,\alpha'}_s)_{s\in[0,t]} \in C_{[0,t]}(B)$ verifies the restrictions
$\phi_0 \equiv \alpha$, $\phi_t \equiv \alpha'$,  it is a minimizer of the variational problem 
in the left-hand side of \eqref{MFreduction}.

The derivation of \eqref{Pequation} follows in the same way as for the Curie-Weiss  model 
in \cite[Section 2.1]{FedHoMa13}, with the Fr\'echet derivative replacing the standard 
derivative. Note that $\alpha \mapsto C_{t,\alpha'}(\alpha)$ is Fr\'echet differentiable on 
$\mathrm{int}(B)$, while the argument in Ellis~\cite[Section V, Theorem $5.1$]{EE83} 
shows that all its critical points lie in $\mathrm{int}(B)$.
\end{proof}

%%%%%%
The following way of rewriting $C_{t,\alpha'}$ will be useful later on.
Adding and subtracting $\tfrac{1}{4} \beta \int_{\T^d} du \int_{\T^d} dv\, J(u-v) [\alpha(u)-\alpha(v)]^2$, 
we may rewrite \eqref{Srf} as
\be
I_S(\alpha)= \tfrac{1}{4} \beta \int_{\T^d} du \int_{\T^d} dv\, J(u-v) [\alpha(u)-\alpha(v)]^2 
+ \int_{\T^d} du\,[-\tfrac{1}{2} \beta \alpha(u)^2 -c \alpha(u)].
\ee
With this formula, \eqref{simplificationcost} reduces to 
\be
\label{PT}
C_{t,\alpha'}(\alpha) = \tfrac{1}{4} \beta \int_{\T^d} du \int_{\T^d} dv\,J(u-v) [\alpha(u)-\alpha(v)]^2
+ \int_{\T^d} du\, C_{t,\alpha'}^\CW(\alpha(u)).
\ee
This form clarifies the interplay between the non-local interaction and the independent spin-flip dynamics.

\subsection{Proof of Theorem \ref{nGBi}}
\label{S2.3}

As emphasized in \eqref{spec-initialm}, $\gamma^{u}_t(k'\mid \alpha')$ depends on 
$\alpha'$ only through $\hat{\phi}^{t,\alpha'}_0$, the starting value of the global minimizer 
of $C_{t,\alpha'}$. The following lemma is the basis for the proof of Theorem~\ref{nGBi}. 
It describes the behavior of $\hat{\phi}^{t,\alpha'}_0$ when the constraint $\alpha'\in B$ 
at time $t$ is varied. Loosely speaking, it says that global minimizers are isolated, are 
continuous under variations of $\alpha'$, and can be selected by variation of $\alpha'$.

Below we fix $t$ and suppress it from the notation. In what follows we write $\hat{\alpha}
(\alpha')$ to denote a global minimum of $C_{t,\alpha'}$.

\begin{lemma}
\label{minimaprops}
For every $t \geq 0$ and $\alpha'_0 \in B$ there exists an open neighborhood 
$\mathcal{N}_{\alpha'_0}$ of $\alpha'_0$ such that for all $\alpha' \in 
\mathcal{N}_{\alpha'_0} \setminus \{\alpha'_0\}$ the following hold:\\
(a) {\rm \bf [Isolation of global minimizers]}
$\alpha \mapsto C_{t,\alpha'}$ has a unique global minimum at, say, $\hat{\alpha}
(\alpha')$.\\
(b) {\rm \bf [Continuity of global minimizers]} 
$\alpha'' \mapsto \hat{\alpha}(\alpha'')$ is continuous at $\alpha''=\alpha'$. If 
$\alpha'' \mapsto C_{t,\alpha'_0}(\alpha'')$ has a unique global minimum, then it is 
continuous at $\alpha''=\alpha'_0$.\\
(c) {\rm \bf [Selection of global minimizers]} 
If $C_{t,\alpha'_0}$ has multiple global minima, then there are two of them, say
$\hat{\alpha}_k(\alpha'_0)$ and $\hat{\alpha}_l(\alpha'_0)$, and a $\gamma' \in B$ 
such that
\be
\label{gamma}
\lim_{\varepsilon \downarrow 0} 
\hat{\alpha}(\alpha'_0+ \varepsilon \gamma') \equiv \hat{\alpha}_k(\alpha'_0),
\qquad 
\lim_{\varepsilon \uparrow 0} 
\hat{\alpha}(\alpha'_0+\varepsilon \gamma')\equiv \hat{\alpha}_l(\alpha'_0).
\ee
\end{lemma}

\bpr 
The following 3 steps describe the behavior of the minimizers under small perturbations 
of $\alpha'$ are around $\alpha'_0$.

\medskip\noindent
(a) 
Under the assumption that $\sup_{\alpha \in B} |C_{t,\alpha'}-C_{t,\alpha_0'}| \to 0$ as $\|\alpha'-\alpha_0'\|_\infty \to 0$, whenever a local minimum is emerging as $\alpha'$ is varied this local minimum cannot be a global minimum. 
Indeed, we have that
$$|C_{t,\alpha'}(\alpha)-C_{t,\alpha'_0}(\alpha)|
\leq 
\int_{\T^d} du \, |C^{CW}_{t,\alpha'(u)}(\alpha(u))-C^{CW}_{t,\alpha'_0(u)}(\alpha(u))| 
\leq
\int_{\T^d} du \, \|C^{CW}_{t,\alpha'(u)}-C^{CW}_{t,\alpha'_0(u)}\|_\infty.
$$
On the other hand, we know from \cite{FedHoMa13} that $\|C^{CW}_{t,m'}-C^{CW}_{t,m'_0}\|_\infty
\to 0$ when $m'\to m'_0$. Hence the claim follows by dominated convergence.

\medskip\noindent
(b) 
Let $\hat{\alpha}_i(\alpha'_0)$, $i\in\cI$, denote the global minima of $C_{t,\alpha'_0}$. 
Each of these verifies \eqref{Pequation}, which may be written in the form $F(\alpha,\alpha') 
\equiv 0$ for some functional $F$. From the implicit function theorem (see e.g.\ Dr\'abek 
and Milota~\cite[Theorem 4.2.1]{DrMi07}) it follows that there exist a neighborhood 
$\widetilde{\mathcal{N}}_{\alpha'_0}$ of $\alpha_0'$ and smooth functions $\alpha' \mapsto 
\overline{\alpha}_i(\alpha')$, $i\in\cI$, on this neighborhood such that $\overline{\alpha}_i
(\alpha')$, $i\in\cI$, are minima of $C_{t,\alpha'}$, and $\lim_{\alpha'\to\alpha'_0}
\overline{\alpha}_i( \alpha') \equiv \hat{\alpha}_i( \alpha'_0)$.

\medskip\noindent
(c) 
Let
\be
B_i(\alpha'):= C_{t,\alpha'}(\overline{\alpha}_i(\alpha')).
\ee
The minimal cost is
\be 
C_{t,\alpha'}(\hat{\alpha}( \alpha'))=\min_{i\in\cI} B_i(\alpha').
\ee
Because of the assumed multiplicity of minima at $\alpha'_0$, we have
\be
\label{eq:bb0}
B_i(\alpha'_0) = B_j(\alpha'_0), \qquad  i,j\in\cI.
\ee
Expand each $B_i$ up to first order order, 
\be
\label{eq:bb1}
B_i(\alpha'_0 + \varepsilon \gamma') = B(\alpha'_0) 
+ \varepsilon \big\langle [DB_i](\alpha'_0), \gamma'\big\rangle 
+ O\big(\varepsilon \|\gamma'\|_\infty\big), \qquad \varepsilon>0,
\ee
where $[DB_i](\alpha'_0)$ is the Fr\'echet derivative. Put $G(\alpha,\alpha')
:=C_{t,\alpha'}(\alpha)$. Then the chain rule implies that
\be
\label{eq:bb2}
[DB_i](\alpha'_0)
\equiv 
[D_\alpha G]\big(\hat{\alpha}_i(\alpha'_0),\alpha'_0\big)
\circ
[D_{\alpha'}\overline{\alpha}_i](\alpha'_0)
+ 
[D_{\alpha'} G]\big(\hat{\alpha}_i(\alpha'_0),\alpha'_0\big),
\ee
where $\circ$ denotes composition and the lower indices $\alpha,\alpha'$ on the letter 
$D$ refer to the variable with respect to which the derivative is taken. The first term 
in \eqref{eq:bb2} vanishes due to the criticality of $\hat{\alpha}_i(\alpha'_0)$. Standard 
calculations with Fr\'echet derivatives show that
\be
\label{eq:bb3}
[D_{\alpha'} G]\big(\hat{\alpha}_i(\alpha'_0), \alpha'_0\big)(u)
=H^{\CW}\big(\hat{\alpha}_i(\alpha'_0)(u), \alpha'_0(u)\big), 
\quad u \in \T^d,
\ee
with $H^{\CW}(m,m'):=(\frac{\partial}{\partial m'}C^{\CW}_{t,m'})(m)$. The identity in 
\eqref{eq:bb3} helps us to select different global minimizers by small variations of 
$\alpha'$. Indeed, for $i\neq j$ we have $\| \hat{\alpha}_i(\alpha'_0)-\hat{\alpha}_j
(\alpha'_0)\|_\infty>0$, and hence there exists a $\delta>0$ such that $\lambda(
\{\hat{\alpha}_i(\alpha'_0)-\hat{\alpha}_j(\alpha'_0)>\delta\})>0$. 
Take $I= \{u \in \T^d \ : \ \hat{\alpha}_i(\alpha'_0)(u)-\hat{\alpha}_j(\alpha'_0)(u)>\delta\}$. Then 
\be
\label{eq:bb4}
\hat{\alpha}_j(\alpha'_0)(u)+ \delta< \hat{\alpha}_i(\alpha'_0)(u) \quad \forall \ u \in I.
\ee
Combining (\ref{eq:bb2}--\ref{eq:bb4}) and using the strict monotonicity of $m \mapsto 
H^{\CW}(m,m')$, we get
\be
[DB_j](\alpha'_0)(u)< [DB_i](\alpha'_0)(u) \quad \forall \ u \in I.
\ee
The claim follows by picking $\gamma'\equiv 1_{I}$ and expressions \eqref{eq:bb1}, \eqref{eq:bb3}.
\epr

We are now ready to prove Theorem~\ref{nGBi}. We continue to use the same notation 
as in Lemma~\ref{minimaprops}.

\bpr
Suppose that $C_{t,\alpha'_0}$ has a unique global minimizer, say $\hat{\alpha}(\alpha'_0)$, 
and let $\mathcal{N}_{\alpha'_0}$ be the neighborhood in Lemma~\ref{minimaprops}. Then 
\eqref{spec-initialm} holds for every $\alpha'\in\cN_{\alpha'_0}$, and the continuity of 
$m \mapsto \Gamma_t(k',m)$ for all $t,k'$ gives the desired continuity of $\alpha' \mapsto 
\gamma^u_t(\cdot \mid \alpha')$ at $\alpha'\equiv\alpha'_0$ for all $u \in \T^d$. Hence 
$\alpha'_0$ is a good profile.

Conversely, suppose that $C_{t,\alpha'_0}$ has multiple global minimizers. 
Consider the pair $\hat{\alpha}_k(\alpha'_0)$ and $\hat{\alpha}_l(\alpha'_0)$ and the box $I$ in the proof of Lemma~\ref{minimaprops}, and put $\alpha'^k_\epsilon:=\alpha'_0 + \epsilon \gamma'$ for $\epsilon>0$ and $\alpha'^l_\epsilon:=\alpha'_0 + \epsilon \gamma'$ for $\epsilon<0$. 
Then $\gamma^u_t(\cdot \mid \alpha'^i_\epsilon) =\Gamma_t (\cdot,\beta\{J \ast \hat{\alpha}
(\alpha'^i_\epsilon)+ h\}(u))$, $i \in \{k,l\}$, and
\be
\lim_{\epsilon \downarrow 0} \hat{\alpha}(\alpha'^k_\epsilon)(u) 
= \hat{\alpha}_k(\alpha'_0)(u) 
\neq \hat{\alpha}_l(\alpha'_0)(u) = \lim_{\epsilon \uparrow 0} 
\hat{\alpha}(\alpha'^l_\epsilon)(u)  \quad \forall \ u \in I. 
\ee
On the other hand, $\hat{\alpha}_k(\alpha'_0)$ and $\hat{\alpha}_l(\alpha'_0)$ are 
critical points, they satisfy \eqref{Pequation} with $\alpha'\equiv \alpha'_0$, and 
so
\be
\hat{\alpha}_k(u)\neq \hat{\alpha}_l(u) \quad \Longrightarrow \quad
(J \ast \hat{\alpha}_k)(u)  \neq (J \ast \hat{\alpha}_l)(u).
\ee
This, together with the continuity and the monotonicity of $m\mapsto \Gamma_t(k',m)$ for all $t$ and $k'$, forces the discontinuity
\be
\begin{aligned}
&\lim_{\epsilon \downarrow 0} \gamma^u_t(k' \mid \alpha'^k_\epsilon) 
= \Gamma_t\big(k',\beta\{J\ast\hat{\alpha}_k(\alpha'_0)+h\}(u)\big)\\
&\qquad \neq \Gamma_t\big(k',\beta\{J\ast\hat{\alpha}_l(\alpha'_0)+h\}(u)\big)
=\lim_{\epsilon \uparrow 0} \gamma^u_t(k' \mid \alpha'^l_\epsilon) 
\quad \forall\,u \in I.
\end{aligned}
\ee
Hence $\alpha'_0$ is a bad profile.
\epr

%%%%%%%%%% SECTION 3 %%%%%%%%%%%%%%%%%%%%%%%%%%%%

\section{Proof of Theorem \ref{Bifurc}}
\label{S3}

\bpr
Without loss of generality we may assume that $\langle J \rangle=1$. For simplicity, 
we consider only $\alpha'\in C(\T^d)$. In that case, due to the regularization property 
of the convolution operator, the solutions of \eqref{Pequation} may be taken to be 
continuous, and \eqref{Pequation} must be fulfilled for all $u \in \T^d$. The extension
to $\alpha'\notin C(\T^d)$ is straightforward. 

\medskip\noindent
(i) Let $\alpha_1, \alpha_2 \in B$ be two different solutions of \eqref{cvp}. After some 
algebra with trigonometrical identities, we get from \eqref{cvp} that the following equation 
must be fulfilled:
\be
\label{uniqueness}
\tfrac{2\sinh\left( \tfrac{A_u-B_u}{2} \right)}{a_u-b_u} 
\Big\{\cosh\left( \tfrac{A_u+B_u}{2} \right)
-a_u \, \sinh\left( \tfrac{A_u+B_u}{2} \right)\Big\} 
- \cosh\left(B_u \right)=\coth(2t) \qquad \forall \ u \in \T^d,
\ee
where $A_u=(\beta J\ast\alpha_1)(u) + \beta h(u)$ and $a_u=\alpha_1(u)$ (and similarly 
for $B_u, b_u, \alpha_2$). Note that the left-hand side depends only on $u$ and the right-hand 
side only on $t$, and that $\lim_{t \downarrow 0} \coth(2t)= \infty$. Since $|A_u|,|B_u| \leq 
\beta(1+\|h\|_\infty)$ and $|a_u|,|b_u|\leq 1$, the left-hand side of \eqref{uniqueness} is 
bounded from above by 
\be
\tfrac{2\sinh\left( \tfrac{A_u-B_u}{2} \right)}{a_u-b_u}\,C_1+C_2
\ee
for some constants $C_1,C_2$. By taking $t>0$ small enough, we force $a_u-b_u$ to 
be small for all $u \in \T^d$ (equivalently, $\|\alpha_1-\alpha_2\|_\infty<\delta$). By choosing 
$v_0$ such that $|\alpha_1(v_0)-\alpha_2(v_0)|=V_0$ with $V_0=\max_{u\in\T^d} |\alpha_1(u)
-\alpha_2(u)|$, we get $|A_{v_0}-B_{v_0}|\leq \beta V_0$ which, together with the series 
expansion of $\sinh$, leads to a contradiction.

\medskip\noindent
(ii) From \eqref{PT}, whenever $\alpha'\equiv c'$ we have that
\be
\inf_{\alpha \in B} C_{t,c'}(\alpha) 
\geq \inf_{\alpha \in B} \tfrac{1}{4} \beta \int_{\T^d} du \int_{\T^d} dv\,
J(u-v) [\alpha(u)-\alpha(v)]^2 + \inf_{\alpha \in B} \int_{\T^d} du\, C_{t,c'}^\CW(\alpha(u)).
\ee
Because $J\geq 0$, the minimizers of the first term are the constant profiles. If we take the constant 
of the profile equal to a minimizer of $C^\CW_{t,c'}$, then the second term is also minimal.
\epr

%%%%%%%%% APPENDIX %%%%%%%%%%%%%%%%%%%%%%%%%%%%%%%

\appendix

\section{Proof of Proposition \ref{PQLDPS}}
\label{appA}

%%%%%%%%%%%%%

\subsection{Outline}

In Sections~\ref{Aub}--\ref{Astart} we sketch the proof of the LDP in Proposition~\ref{PQLDPS}(i) 
for deterministic initial conditions (as in Proposition~\ref{LDPCo}(ii)), and explain why it 
remains true for random initial conditions. We follow the line of argument in Benois, 
Mourragui, Orlandi, Saada and Triolo~\cite{BeMoOrSaTr12} rather than Comets~\cite{Co87},
and use various results from Kipnis and Landim~\cite{KipLan99}. The strategy of the proof consists 
in first proving the claim for random initial conditions drawn according to $\vartheta_\kappa^n
=\otimes_{x\in\T_n^d} \vartheta_{\kappa}$ with $\vartheta_{\kappa}=\mathrm{BER}(\kappa)$, 
$\kappa \in [0,1]$ (i.e., $\vartheta_\kappa(+1)=\kappa$ and $\vartheta_\kappa(-1) = 1-\kappa$), 
and afterwards replacing $\vartheta_{\kappa}^n$ by $\mu^n$ in \eqref{Kac} with the help of 
Varadhan's Lemma and Bryc's Lemma. In Section~\ref{ACP} we indicate how 
Proposition~\ref{PQLDPS}(ii) follows. 

Below we will make frequent reference to formulas in \cite{BeMoOrSaTr12} and \cite{KipLan99}, 
so our arguments are not self-contained. We begin with the following observation.

\begin{lemma}
\label{RNlemma}
Suppose that $\mu$ and $\nu$ are equivalent probability measures.  If $P_\mu$ and $Q_\nu$ 
are the laws of equivalent Markov processes with starting measures $\mu$ and $\nu$, then
\be
\label{RN}
\frac{dP_\mu}{dQ_\nu}(\bar{\eta})=\frac{d\mu}{d\nu}(\eta_0) \, \frac{dP_\mu}{dQ_\mu}(\bar{\eta})
=\frac{d\mu}{d\nu}(\eta_0) \, \frac{dP_\nu}{dQ_\nu}(\bar{\eta}).
\ee
\end{lemma}

\noindent
The general technique to prove an LDP relies on finding a family of mean-one positive martingales 
that can be written as functions of the empirical density. For Markov processes this is achieved by 
considering the Radon-Nikodym derivative of the original dynamics w.r.t.\ a small perturbation of 
this dynamics. It is here that Lemma~\ref{RN} comes into play: it factorizes the Radon-Nikodym 
derivative into a \emph{static part} and a \emph{dynamic part}, as in \eqref{SDrf}. 

%%%%%%%%%%%%%%%

\subsection{Upper bound}
\label{Aub}

For initial condition $\gamma \in C(\T^d;[-1,+1])$ and potential $V \in C^{1,0}([0,t]\times\T^d)$, we 
denote by $\mathbb{P}^{n,V}_{\vartheta^n_{\gamma}}$ the law of the $(\gamma,V)$-perturbed 
inhomogeneous Markov process starting at 
\be
\label{startmeas}
\vartheta^n_\gamma=\otimes_{x\in\T_n^d} 
\vartheta_{\chi^{-1}\left(\gamma(\tfrac{x}{n})\right)},
\ee
where $\chi\colon[0,1] \to [-1,+1]$ is the linear map that transforms a profile taking values in $[-1,+1]$ 
into a profile taking values in $[0,1]$. Details about such a perturbation and its Radon-Nikodym 
derivative can be found in \cite[Eq.~(5.8)]{BeMoOrSaTr12}.

\medskip\noindent
{\bf 1.} 
\emph{Large deviation upper bound for compact sets}.
Fix $\kappa\in[0,1]$. Let $\mathcal{K} \in D_{[0,t]}(\M(\T^d))$ be compact. By Lemma~\ref{RNlemma}, 
we have (recall the notation introduced in Section~\ref{S1.5})
\be
\begin{aligned}
\tfrac{1}{n^d} \log \mathbb{P}^n_{\vartheta_\kappa^n}[\bar{\pi}^n \in \mathcal{K}]
&= \tfrac{1}{n^d} \log \mathbb{E}^{n,V}_{\vartheta^n_\gamma}
\left[\left(\frac{d\mathbb{P}^n_{\vartheta_\kappa^n}}{d\mathbb{P}^{n,V}_{\vartheta^n_\gamma}}
\, \If{\mathcal{K}}\right)(\bar{\pi}^n)\right]\\
&= \tfrac{1}{n^d} \log \mathbb{E}^{n,V}_{\vartheta^n_\gamma}
\left[\left(\frac{d\vartheta_\kappa^n}{d\vartheta^n_\gamma} \,
\frac{d\mathbb{P}^n_{\vartheta_\kappa^n}}{d\mathbb{P}^{n,V}_{\vartheta^n_\kappa}} 
\, \If{\mathcal{K}}\right)(\bar{\pi}^n)\right]\\
&= \tfrac{1}{n^d} \log \mathbb{E}^{n,V}_{\vartheta^n_\gamma}
\left[e^{-n^d  h_\gamma(\pi^n_0)+O_\gamma(n^{-1})}
\,e^{-n^d \{\hat{J}_V(\bar{\pi}^n \ast l^{\varepsilon,n}) + r(V,\varepsilon,n)\}}
\, \If{\mathcal{K}}(\bar{\pi}^n)\right],
\end{aligned}
\ee
where $h_\gamma$ is the analogue of \cite[Eq.\ (1.1), Chapter 10]{KipLan99}, $\hat{J}_V$ is 
defined in \cite[Eq.\ (6.8)]{BeMoOrSaTr12}, $\varepsilon>0$ is small, $l^{\varepsilon,n}$ is an 
approximation of the identity for $\varepsilon \downarrow 0$, and $r(V,\varepsilon,n)$ is an 
error term that vanishes as $n\to\infty$ for fixed $V,\varepsilon$. By letting $n \to \infty$, 
optimizing over $\gamma,V,\varepsilon$ and using the mini-max lemma, we get
\be
\begin{aligned}
\limsup_{n \to \infty}\tfrac{1}{n^d} \log \mathbb{P}^n_{\vartheta_\kappa^n}[\bar{\pi}^n \in \mathcal{K}]
& \leq \inf_{\gamma,V,\varepsilon} 
\sup_{\bar{\pi}\in \mathcal{K}}\{-h_\gamma(\pi_0)-\hat{J}_V(\bar{\pi}\ast l^\varepsilon)\}\\
&\leq \sup_{\bar{\pi}\in \mathcal{K}} \inf_{\gamma,V,\varepsilon} \{- h_\gamma(\pi_0)
-\hat{J}_V(\bar{\pi}\ast l^\varepsilon)\}\\
&\leq -\inf_{\bar{\pi}\in \mathcal{K}} \{I_S(\pi_0) + I^t_D(\bar{\pi})\}.
\end{aligned}
\ee
The last inequality uses that $\sup_\gamma h_\gamma(\pi_0)=I_S(\pi_0)$, $\sup_V \hat{J}_V
(\tilde{\pi}) = I^t_D(\tilde{\pi})$, and $\sup_\varepsilon I^t_D(\bar{\pi}\ast l^\varepsilon) \geq
I^t_D(\bar{\pi})$ by lower semi-continuity of $I^t_D$.

\medskip\noindent
{\bf 2.} 
\emph{Exponential tightness}. 
While in \cite[Section 4]{KipLan99} the initial condition is drawn from equilibrium, this is 
immaterial. Indeed, the proof of \cite[Proposition 6.1]{BeMoOrSaTr12} uses the same ideas 
as in \cite[Section 4]{KipLan99} even though the initial condition is deterministic. Hence the 
same computations apply to our case. 

%%%%%%%%%%%%%%%%%%%%

\subsection{Lower bound}
\label{Alb}

{\bf 1.}
\emph{Large deviation lower bound for open sets}. 
Fix $\kappa\in[0,1]$. Let $\mathcal{O} \in D_{[0,t]}(\M(\T^d))$ be open. By Lemma~\ref{RNlemma}, 
we have
\be
\begin{aligned}
\tfrac{1}{n^d} \log \mathbb{P}^n_{\vartheta_\kappa^n}[\bar{\pi}^n \in \mathcal{O}]
&= \tfrac{1}{n^d} \log \left\{\mathbb{E}^{n,V}_{\vartheta^n_\gamma}\,
\left[\frac{d\mathbb{P}^n_{\vartheta_\kappa^n}}{d\mathbb{P}^{n,V}_{\vartheta^n_\gamma}}
(\bar{\pi}^n)~\Bigg|~ \bar{\pi}^n \in \mathcal{O}\right]
\mathbb{P}^{n,V}_{\vartheta^n_\gamma}(\mathcal{O})\right\}\\
&\geq \mathbb{E}^{n,V}_{\vartheta^n_\gamma} \left[\tfrac{1}{n^d} 
\log\frac{d\mathbb{P}^n_{\vartheta_\kappa^n}}{d\mathbb{P}^{n,V}_{\vartheta^n_\gamma}}(\bar{\pi}^n) 
~\Bigg|~ \bar{\pi}^n \in \mathcal{O}\right]
+ \tfrac{1}{n^d} \log \mathbb{P}^{n,V}_{\vartheta^n_\gamma}(\mathcal{O}),
\end{aligned}
\ee
where we use Jensen's inequality. By the law of large numbers for $\mathbb{P}^{n,V}_{\vartheta^n_\gamma}$,
we have 
\begin{equation}
\label{LLNPnV}
w-\lim_{n\to\infty} \mathbb{P}^{n,V}_{\vartheta^n_\gamma} = \delta_{\bar{\pi}^{\gamma, V}},
\end{equation} 
where $\bar{\pi}^{\gamma,V}$ is the solution of \cite[Eq.\ (5.5)]{BeMoOrSaTr12} with initial 
condition $\gamma$ and potential $V$. (The proof of \eqref{LLNPnV} follows in the same 
fashion as in \cite{BeMoOrSaTr12}: all that is needed is that the laws of the random initial 
conditions converge to a law associated with continuous profile.) Hence, if $\bar{\pi}^{\gamma,V} 
\in \mathcal{O}$, then $\lim_{n\to\infty} \mathbb{P}^{n,V}_{\vartheta^n_\gamma}(\mathcal{O})
=1$. After some calculations with the Radon-Nikodym derivative, we get
\be
\liminf_{n \to \infty} \tfrac{1}{n^d} \log \mathbb{P}^n_{\vartheta_\kappa^n}[\bar{\pi}^n \in \mathcal{O}]
\geq -I^t(\bar{\pi}^{\gamma,V})
\ee
with $I^t = I_S + I^t_D$.

\medskip\noindent
{\bf 2.}
\emph{Density arguments}. 
It remains to show that
\be
\inf_{\substack{\gamma,V \\ \bar{\pi}^{\gamma,V} \in \mathcal{O}}} I^t(\bar{\pi}^{\gamma,V})
= \inf_{\bar{\pi} \in \mathcal{O}} I^t(\bar{\pi}).
\ee
In other words, $(\bar{\pi}^{\gamma,V})_{\gamma, V}$ is dense with respect to $(\varrho^w_t, I)$,
i.e.,
\be
\begin{aligned}
&\forall \,\bar{\pi}\in D_{[0,t]}(\M(\T^d))\colon\, I(\bar{\pi})< \infty, \\ 
&\exists \,(\bar{\pi}^{\gamma_n, V_n})_{n\in\N}\colon\, 
\lim_{n\to\infty} \varrho^w_t(\bar{\pi}^{\gamma_n, V_n},\bar{\pi}) = 0, \,
\lim_{n\to\infty} I(\bar{\pi}^{\gamma_n, V_n}) = I(\bar{\pi}),
\end{aligned}
\ee
where $\varrho^w_t$ is the supremum distance in $[0,t]$ when the marginal distance 
is $\varrho^w$ (any metric that metrizes the weak topology). A density argument of this 
type typically exploits the fact that $I$ is lower semi-continuous and convex, but in our 
case $I=I^t$, which is not convex. However, in \cite{BeMoOrSaTr12} density arguments 
are given without convexity. In order to extend these to our setting of random initial 
conditions, minor modifications are needed in \cite[Lemma 7.5]{BeMoOrSaTr12}. 
In particular, the space regularization of the trajectory must be done for all $s \in [0,t]$, and 
hence \cite[Lemma 7.6]{BeMoOrSaTr12} together with the arguments in \cite[p.\ 279]{KipLan99}
prove our assertion.

%%%%%%%%%%%%%%%%%%%%

\subsection{Replace $\vartheta^n_\kappa$ by $\mu^n$}
\label{Astart}

The observations made in Sections~\ref{Aub}--\ref{Alb} prove the LDP in Proposition~\ref{PQLDPS}(i),
but for starting measures $\vartheta^n_\kappa$ given by \eqref{startmeas}. Note that
\be
\frac{d\mu^n}{d \vartheta^n_\kappa}= e^{n^d \beta H(\pi^n)}
\ee
with $\pi^n \mapsto H(\pi^n)$ in \eqref{Hamiltonianrew} continuous. Hence, by Lemma \eqref{RNlemma}, 
Varadhan's Lemma and Bryc's Lemma, the LDP in Proposition~\ref{PQLDPS}(i) for starting measures 
$\mu^n$ follows.

%%%%%%%%%%%%%%%%%%%%%

\subsection{Contraction principle}
\label{ACP}

Proposition~\ref{PQLDPS}(ii) follows from Proposition~\ref{PQLDPS}(i) via the approximate contraction 
principle based on exponential approximation estimates. See Dembo and Zeitouni~\cite[Section 4.2]{DeZe10}.

%%%%%%%%% REFERENCES %%%%%%%%%%%%%%%%%%%%%%%%%%%%%

\end{document}